\documentclass[pra,aps,10pt,superscriptaddress,twocolumn,floatfix]{revtex4-1}
\usepackage{graphicx,color}
\usepackage[nice]{nicefrac}							
\usepackage{amsmath,amssymb,bm}

\usepackage[plainpages=false,pdfpagelabels,colorlinks=true,linkcolor=red,urlcolor=blue,citecolor=blue,pdftitle={},pdfauthor={},pdfdisplaydoctitle=true,pdfduplex=DuplexFlipLongEdge]{hyperref}

\definecolor{darkred}{rgb}{0.90,0.2,0.2}
\definecolor{darkgreen}{rgb}{0,0.60,.2}
\definecolor{darkblue}{rgb}{0.1,0.3,1}
\definecolor{grey}{cmyk}{0,0,0,0.25}
\definecolor{orange}{cmyk}{0,0.6,0.8,0}

\begin{document}

\title{Quantum chaos challenges many-body localization}

\author{Jan \v Suntajs}
\affiliation{Department of Theoretical Physics, J. Stefan Institute, SI-1000 Ljubljana, Slovenia}
\author{Janez Bon\v ca}
\affiliation{Department of Physics, Faculty of Mathematics and Physics, University of Ljubljana, SI-1000 Ljubljana, Slovenia}
\affiliation{Department of Theoretical Physics, J. Stefan Institute, SI-1000 Ljubljana, Slovenia}
\author{Toma\v z Prosen}
\affiliation{Department of Physics, Faculty of Mathematics and Physics, University of Ljubljana, SI-1000 Ljubljana, Slovenia}
\author{Lev Vidmar}
\affiliation{Department of Theoretical Physics, J. Stefan Institute, SI-1000 Ljubljana, Slovenia}
\affiliation{Department of Physics, Faculty of Mathematics and Physics, University of Ljubljana, SI-1000 Ljubljana, Slovenia}

\begin{abstract}
Characterizing states of matter through the lens of their ergodic properties is a fascinating new direction of research.
In the quantum realm, the many-body localization (MBL) was proposed to be the paradigmatic ergodicity breaking phenomenon, which extends the concept of Anderson localization to interacting systems.
At the same time, random matrix theory has established a powerful framework for characterizing the onset of quantum chaos and ergodicity (or the absence thereof) in quantum many-body systems.
Here we numerically study the spectral statistics of disordered interacting spin chains, which represent prototype models expected to exhibit MBL.
We study the ergodicity indicator $g=\log_{10}(t_{\rm H}/t_{\rm Th})$, which is defined through the ratio of two characteristic many-body time scales, the Thouless time $t_{\rm Th}$ and the Heisenberg time $t_{\rm H}$, and hence resembles the logarithm of the dimensionless conductance introduced in the context of Anderson localization.
We argue that the ergodicity breaking transition in interacting spin chains occurs when both time scales are of the same order, $t_{\rm Th} \approx t_{\rm H}$, and $g$ becomes a system-size independent constant.
Hence, the ergodicity breaking transition in many-body systems carries certain analogies with the Anderson localization transition.
Intriguingly, using a Berezinskii-Kosterlitz-Thouless correlation length we observe a scaling solution of $g$ across the transition, which allows for detection of the crossing point in finite systems.
We discuss the observation that scaled results in finite systems by increasing the system size exhibit a flow towards the quantum chaotic regime.
\end{abstract}

\maketitle

Quantum many-body physics is currently facing a revival in addressing long-standing open questions, such as the emergence of quantum ergodicity in nonequilibrium systems of interacting particles.
An extensive amount of recent theoretical and experimental work~\cite{eisert_friesdorf_15, dalessio_kafri_16, mori_ikeda_18, Trotzky2012, Kaufman2016, Neill2016} established a view that generic (disorder free) quantum many-body systems can usually be considered as quantum ergodic, suggesting that local observables thermalize after the system has been driven away from equilibrium.

A remarkable phenomenon where ergodicity is absent is the Anderson localization~\cite{anderson_58}.
One of the cornerstones of the latter is the scaling solution of the dimensionless conductance~\cite{abrahams_anderson_79}, which sheds light onto the Anderson localization transition in different dimensions~\cite{lee_ramakrishnan_85, kramer_mackinnon_93}.
One commonly expresses the dimensionless conductance as the ratio $\tilde t_{\rm Th}/\tilde t_{\rm H}$, which describes the scaling properties of the characteristic  time denoted the single-particle Thouless time ($\tilde t_{\rm Th}$) in units of the single-particle Heisenberg time ($\tilde t_{\rm H}$), the latter being proportional to the inverse level spacing of the single-particle spectrum.
The time $\tilde t_{\rm Th}$ can be viewed as an inverse of the energy scale that probes sensitivity of disordered systems to boundary conditions (as introduced by Edwards and Thouless~\cite{edwards_thouless_72}), and it is now commonly referred to as the time after which the quantum dynamics is universal and governed by random matrices.
Following the latter definition of $\tilde t_{\rm Th}$, it has been recently shown that requiring $\tilde t_{\rm Th}/\tilde t_{\rm H} = {\rm const}$ when increasing the system size provides a very efficient tool to detect the localization transition point in the three and five dimensional Anderson models~\cite{sierant_delande_20}.

The ideas to extend the framework of localization transitions to interacting systems gave rise to the possibility of a new class of phase transitions denominated the MBL transition~\cite{pal_huse_10} (see also~\cite{basko_aleiner_06, gornyi_mirlin_05, oganesyan_huse_07, imbrie_16_short, Rahul15, altman_vosk_15, abanin_altman_19}).
Despite the considerable ongoing efforts, universal properties of the MBL transition are still intensively debated.
From the perspective of exact numerical solutions, there is still lack of consensus about the position and the nature of the transition point in paradigmatic models (in particular, disordered spin-1/2 chains, our focus here)~\cite{oganesyan_huse_07, pal_huse_10, monthus_garel_10, barkelbach_reichman_10, Luitz2015, devakul_singh_15, bera_schomerus_15, bertrand_garciagarcia_16, khemani_lim_17, doggen_schindler_18, gray_bose_18, mace_alet_19, roy_chalker_19, abanin_bardarson_20, suntajs_bonca_20}.
Moreover, phenomenological theories such as the real-space renormalization group are also experiencing rapid development~\cite{vosk_huse_15, potter_vasseur_15, zhang_zhao_16, dumitrescu_vasseur_17, dumitrescu_goremykina_19, goremykina_vasseur_19, morningstar_huse_19, morningstar_huse_20}, with the current understanding that the transition exhibits some properties of the Berezinskii-Kosterlitz-Thouless transition~\cite{dumitrescu_goremykina_19, goremykina_vasseur_19, morningstar_huse_19, morningstar_huse_20}.

The MBL transition is conjectured to be an eigenstate phase transition from an ergodic to a nonergodic phase~\cite{pal_huse_10, bauer_nayak_13, huse_nandkishore_13, pekker_refael_14}.
Signatures of the transition are usually explored by studying the structure of Hamiltonian eigenstates~\cite{pal_huse_10, bauer_nayak_13, kjall_bardarson_14} and matrix elements of local observables~\cite{pal_huse_10, khatami_rigol_12, serbyn_papic_15, serbyn_papic_17, panda_scardicchio_20}.
Another convenient approach to characterize the MBL transition, which is inspired by successful analyses of the Anderson localization transition~\cite{shklovskii_shapiro_93, zharekeshev_kramer_97}, is based on studies of spectral properties~\cite{oganesyan_huse_07, serbyn_moore_16, buijsman_cheianov_19, sierant_zakrzewski_19, suntajs_bonca_20}.
The key reference point for the latter is provided by the random matrix theory and the so-called quantum chaos conjecture~\cite{casati_valzgris_80, bohigas_giannoni_84, berry_85}.
As a consequence, the emergence of random-matrix-theory-like spectral properties in isolated many-body quantum systems is commonly considered as a working definition of the many-body quantum chaos~\cite{dalessio_kafri_16}.

A standard approach pursued so far in the context of disordered many-body systems (with some exceptions~\cite{bertrand_garciagarcia_16, serbyn_moore_16, sierant_zakrzewski_19, schiulaz_torresherrera_19}) is to study spectral properties by means of level spacing analyses.
The mean level spacing $\overline{\delta E}$ in those systems typically decays exponentially with the number of lattice sites $L$, and hence its inverse, called the (many-body) Heisenberg time $t_{\rm H} = \hbar/\overline{\delta E}$, increases exponentially with $L$.
However, much more comprehensive information is obtained by studying spectral properties on all energy (and time) scales.
Indeed, such an approach has been recently proved useful in studies of models relevant for the holographic duality~\cite{cotler_gurari_17, cotler_hunterjones_17, gharibyan_hanada_18, xiao_ludwig_18} and in Floquet systems~\cite{kos_ljubotina_18, bertini_kos_18, chan_deluca_18a, chan_deluca_18b}.
One of the key quantities that characterizes the global spectral properties is the Thouless time $t_{\rm Th}$ (or, equivalently, the Thouless energy $E_{\rm Th} \equiv \hbar/t_{\rm Th}$), which can be viewed as the many-body analog of the single particle Thouless time $\tilde t_{\rm Th}$ introduced above.
In diffusive systems $t_{\rm Th}$ is expected to increase with lattice size as $t_{\rm Th} \propto L^2$~\cite{dalessio_kafri_16}, and it can also be referred to as the ergodisation time since it detects the onset of universal quantum chaotic dynamics.

In this work we introduce an indicator of the ergodicity breaking transition in finite many-body systems,
\begin{equation} \label{def_g}
g=\log_{10}(t_{\rm H}/t_{\rm Th}) \,,
\end{equation}
which is proportional to the logarithm of the dimensionless conductance $t_{\rm Th}/t_{\rm H}$ and hence carries similarities with the dimensionless conductance in studies of Anderson localization.
The ergodicity indicator $g$ interpolates between the quantum ergodic regime $t_{\rm Th}/t_{\rm H} \to 0$ [$g \to \infty$] and the nonergodic regime $t_{\rm Th}/t_{\rm H} \to \infty$ [$g \to -\infty$] in the thermodynamic limit.
Our main result is that $g$ exhibits a scaling solution across the ergodicity breaking transition in certain disordered spin chains at $t_{\rm Th} \approx t_{\rm H}$.
The scaling solution exhibits clear signatures of a crossing point that can be readily detected in rather small systems.
The finite-size analysis indicates robustness of quantum chaos upon increasing the disorder; however, limitations to relatively small lattice sizes also prevent us from making conclusions about the fate of the critical regime in the thermodynamic limit.

\begin{figure}[!b]
\begin{center}
\includegraphics[width=0.99\columnwidth]{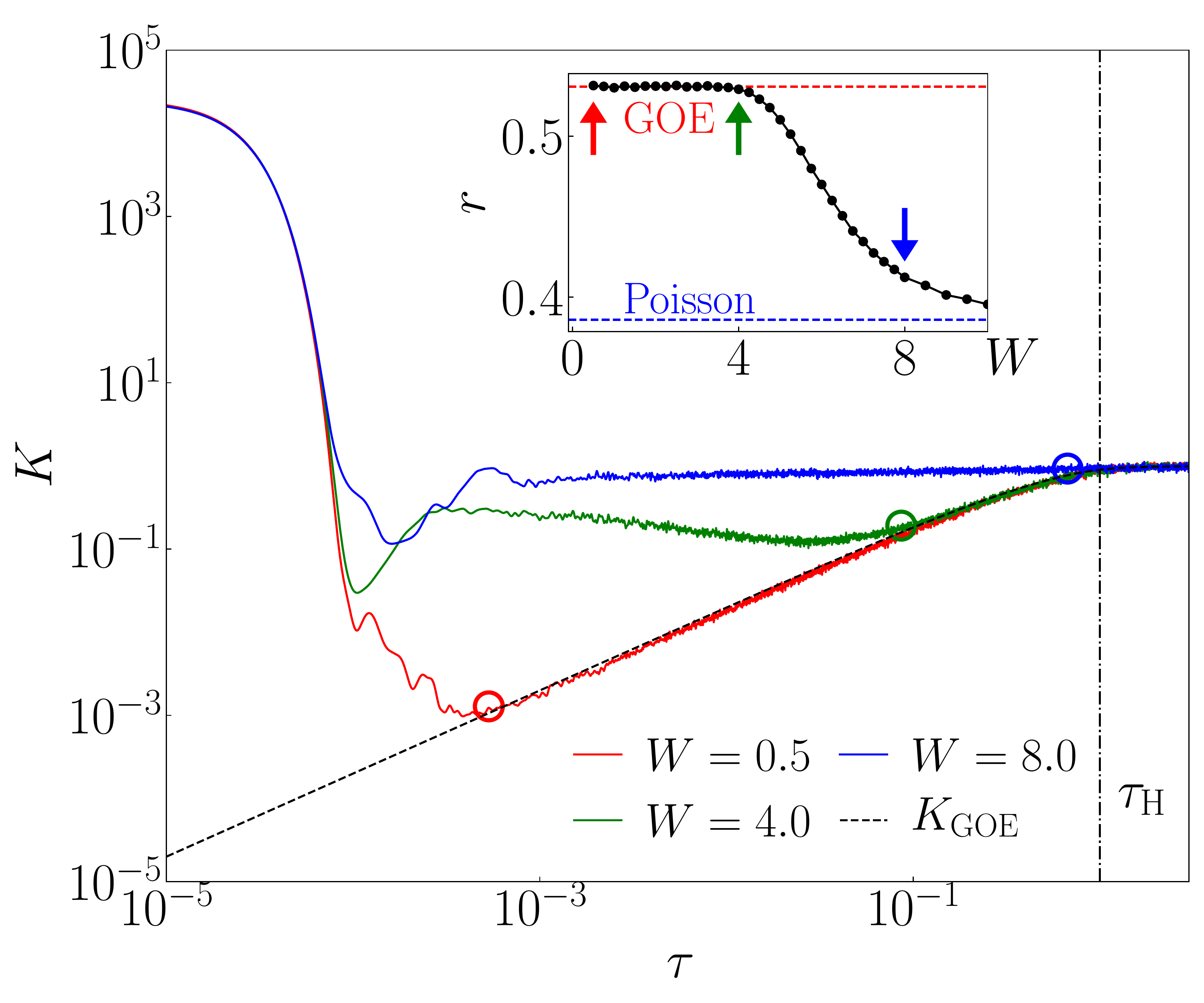}
\end{center}
\caption{
{\it Spectral form factor $K(\tau)$ in the $J_1$-$J_2$ model.}
Results are shown for three disorder strengths $W=0.5, 4, 8$ (lower, middle and upper lines, respectively) on a lattice with $L=18$ sites.
The dashed line is the Gaussian orthogonal ensemble (GOE) prediction $K_{\rm GOE}(\tau)$.
The spectral form factor is a function of the scaled time $\tau$, which is related to time in physical units $t$ through the spectral unfolding procedure.
Vertical line denotes the scaled Heisenberg time $\tau_{\rm H}\equiv1$, while open circles denote the scaled Thouless time $\tau_{\rm Th}$ obtained by using the criterion $\log_{10}[K(\tau_{\rm Th})/K_{\rm GOE}(\tau_{\rm Th})] = 0.08$.
While $K(\tau)$ takes into account spectral correlations at all energy scales, in the inset we also show results for the average level spacing ratio $r$, which takes into account only the nearest energy levels and is a universal constant $r = r_{\rm GOE} = 0.5307$~\cite{atas_bogomolny_13} if $\tau_{\rm Th}/\tau_{\rm H} = t_{\rm Th}/t_{\rm H} \ll 1$.
The arrows in the inset are located at values of $W$ used in the main panel.
See Appendixes~\ref{sec:def_tH}-\ref{sec:def_r} for the definition of all quantities in the figure.
\label{fig1} }
\end{figure}

\begin{figure*}
\begin{center}
\includegraphics[width=2.08\columnwidth]{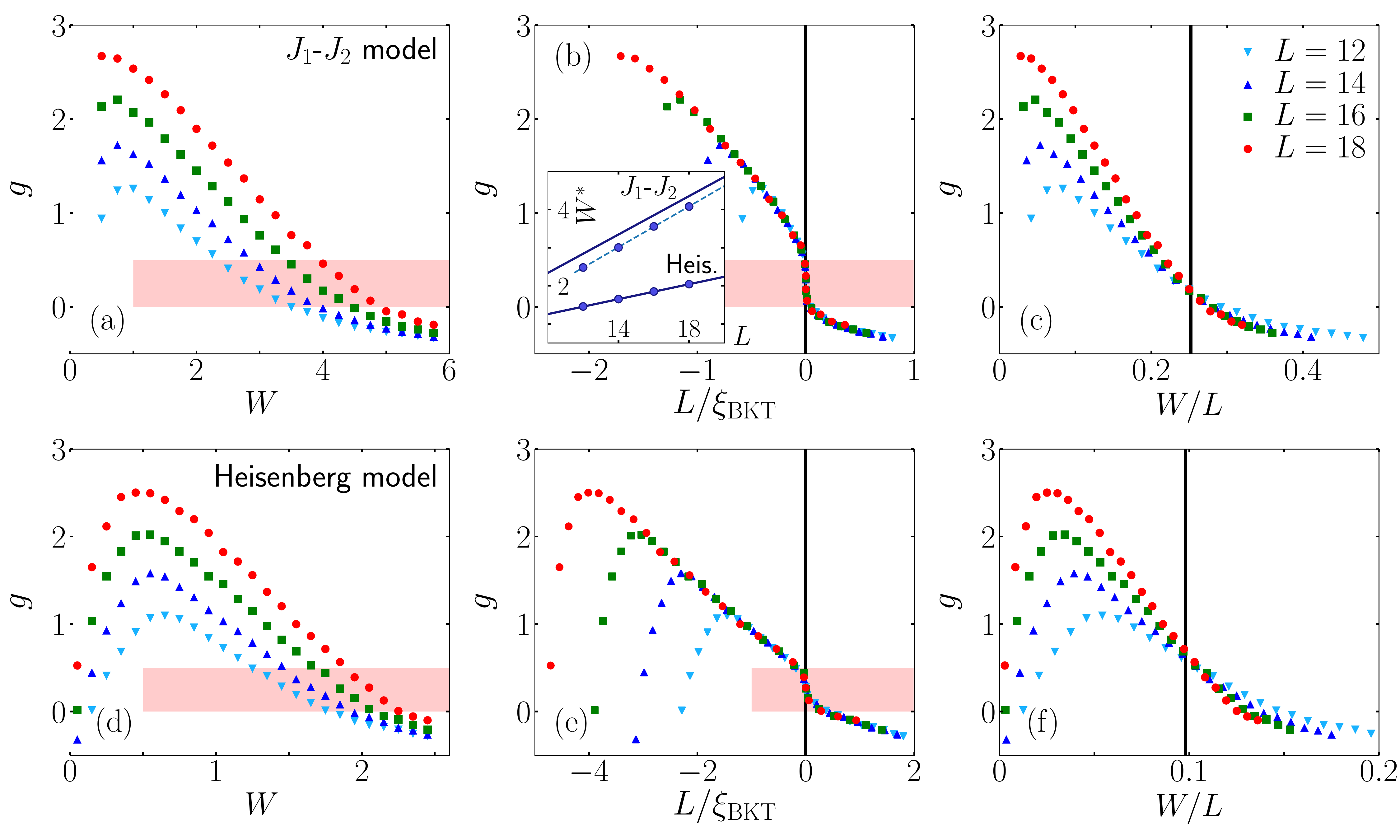}
\end{center}
\caption{
{\it Ergodicity indicator $g$ in disordered spin chains.}
Results are shown for the $J_1$-$J_2$ model (upper panels) and for the Heisenberg model (lower panels).
(a) and (d): $g$ versus disorder $W$ for different lattice sizes $L$, as indicated in the legend.
(b) and (e): $g$ versus $L/\xi_{\rm BKT}$ using the cost function minimization algorithm (see Appendix~\ref{sec:def_cost} and Ref.~\cite{suntajs_bonca_20} for details) with the BKT correlation length $\xi_{\rm BKT}$ from Eq.~(\ref{def_xi_kt}) and the crossing point fitting function $W^* = w_0 + w_1 L$.
In (b) and (e) we set $\xi_{\rm BKT} \to$ sign[$W$-$W^*$] $\xi_{\rm BKT}$.
In the inset of (b), we compare the fitting function $W^*= w_0 + w_1 L$ [solid lines] to a general ansatz for the transition point $W^* = w^*(L)$ [symbols], where independent fitting values for each $L$ are used, yielding rather accurate agreement.
Dashed line is a linear fit to $w^*(L)$ for the $J_1$-$J_2$ model, exhibiting nearly identical slope as the solid line.
Shaded areas in (a)-(b) and (d)-(e) mark the interval of $g$ at which the sharp drop appears at $L/\xi_{\rm BKT}  \approx 0$.
(c) and (f): $g$ versus $W/L$.
Vertical lines show $w_1$ from the optimal crossing point fitting function $W^*= w_0 + w_1 L$ used in (b) and (e).
\label{fig2} }
\end{figure*}

{\bf Disordered spin-1/2 chains.}
A paradigmatic class of Hamiltonians that are expected to exhibit MBL are disordered spin chains with local interactions and on-site disorder,
\begin{align} \label{def_Ham}
 \hat H & =  J_1 \sum_{\ell=1}^{L} \left( \hat s_{\ell}^x \hat s_{\ell+1}^x + \hat s_{\ell}^y \hat s_{\ell+1}^y + \Delta \hat s_{\ell}^z \hat s_{\ell+1}^z \right) \\
 + J_2 & \sum_{\ell=1}^{L} \left( \hat s_{\ell}^x \hat s_{\ell+2}^x + \hat s_{\ell}^y \hat s_{\ell+2}^y + \Delta \hat s_{\ell}^z \hat s_{\ell+2}^z \right) + \sum_{\ell=1}^L w_{\ell} \hat s_\ell^z \, , \nonumber
\end{align}
where $\hat s_\ell^\alpha$ ($\alpha = x,y,z$) are spin-1/2 operators at site $\ell$ and $L$ is the number of lattice sites.
This class of Hamiltonians contains two local conserved operators, the total energy $\hat H$ and the total spin magnetization $\hat s^z = \sum_\ell \hat s_\ell^z$.
Throughout the paper we focus on the $s^z=0$ symmetry sector, and we use periodic boundary conditions, $\hat s^\alpha_\ell = \hat s^\alpha_{L+\ell}$.
Disorder is introduced by independent and identically distributed local magnetic fields, with values $w_\ell/|J_1| \in [-W,W]$ drawn from a uniform distribution, and hence we refer to $W$ as the disorder strength.
We study two models with disorder, the isotropic Heisenberg model [$J_1=1$, $J_2=0$  and $\Delta = 1$ in Eq.~(\ref{def_Ham})] and the anisotropic $J_1$-$J_2$ model [$J_1=J_2 =-1$ and $\Delta=-0.55$ in Eq.~(\ref{def_Ham})].

We use one of the most universal tools for calculating $t_{\rm Th}$, i.e., a Fourier transform of the spectral two-point correlations, known as the spectral form factor $K(\tau)$.
[See Appendixes~\ref{sec:def_tH} and~\ref{sec:def_tTh} for the definition of $K(\tau)$ and for details about the protocols applied to extract $t_{\rm Th}$ and $t_{\rm H}$ in numerical simulations].
Examples of the spectral form factor $K(\tau)$ in the $J_1$-$J_2$ model at weak ($W$=0.5), moderate ($W$=4) and strong ($W$=8) disorder are shown in Fig.~\ref{fig1}.
The spectral form factor $K(\tau)$ represents an insightful measure of the onset of ergodicity in quantum dynamics, which may be a relevant observable for future experiments with analog quantum simulators that study disordered many-body systems~\cite{schreiber_hodgman_15, kondov_mcgehee_15, smith_lee_16, choi_hild_16, lukin_rispoli_19, rispoli_lukin_19, chiaro_neill_20}.
In particular, an experimental protocol has been recently proposed to measure $K(\tau)$ using a quantum non-demolition coupling via an auxiliary 'clock' qubit in Rydberg atom systems~\cite{vasilyev_grankin_20}.

\newpage
{\bf Scaling solution across the transition.}
Numerical results for the ergodicity indicator $g$ as a function of the disorder $W$ and the lattice size $L$ are shown in Figs.~\ref{fig2}(a) and~\ref{fig2}(d) for both models under investigation.
They exhibit two main features:
(i) $g$ increases with $L$ in the disorder regime under investigation, and
(ii) for a fixed $L$, $g$ decreases with increasing $W$ [except for the limit $W \to 0$ in Fig.~\ref{fig2}(d), which is a fingerprint of integrability of the clean Heisenberg chain].
It is therefore far from obvious what is the behavior of $g$ when both $L$ and $W$ are large.
As a technical remark, we note that the protocol to extract $t_{\rm Th}$ from the spectral form factor (see Appendix~\ref{sec:protocol}) is accurate up to a small global multiplicative factor, and hence $g(W,L)$ is determined up to a constant shift.
Determining the exact numerical value of $g$ at the transition is beyond the scope of this work.

The main focus of our analysis are the scaling properties of $g(W,L)$ across the transition.
We explore the scaling as a function of $L/\xi_{\rm BKT}$, where
\begin{equation} \label{def_xi_kt}
 \xi_{\rm BKT} = e^{\frac{b}{\sqrt{|W-W^*|}}}
\end{equation}
is the correlation length of the Berezinskii-Kosterlitz-Thouless (BKT) type, $b$ is a constant, and $W^*$ is the critical point.
In finite systems, we denote $W^* = W^*(L)$ as a crossing point and we relax the condition for $W^*$ to be an $L$-independent number.
The relevance of the BKT correlation length for the MBL transition was identified in the renormalization group calculations~\cite{dumitrescu_goremykina_19, goremykina_vasseur_19, morningstar_huse_19} based on the quantum avalanche picture~\cite{deroeck_huveneers_17, thiery_huveneers_18}.
Exact numerical calculations, however, seemed to indicate~\cite{kjall_bardarson_14, Luitz2015, bertrand_garciagarcia_16, khemani_sheng_17} that the optimal scaling collapse of ergodicity indicators is obtained by using the power-law correlation length $\xi_0 = |W - W^*|^{-\nu}$.
Nevertheless, it was recently argued that the BKT correlation length $\xi_{\rm BKT}$ yields a more favorable scaling solution than $\xi_0$~\cite{suntajs_bonca_20}.

Using the correlation length $\xi_{\rm BKT}$ from Eq.~(\ref{def_xi_kt}) we find an excellent scaling collapse of $g$ versus $L/\xi_{\rm BKT}$ as shown in Figs.~\ref{fig2}(b) and~\ref{fig2}(e).
The scaled function exhibits a sharp drop at $L/\xi_{\rm BKT} \approx 0$, which pins the crossing point in finite systems.
The values of $g$ for which $L/\xi_{\rm BKT} \approx 0$ are marked by the shaded regions in Figs.~\ref{fig2}(a)-\ref{fig2}(b) and~\ref{fig2}(d)-\ref{fig2}(e).
Using the cost function minimization approach~\cite{suntajs_bonca_20} we quantitatively verified that the scaling solution for $g$ as a function of $L/\xi_{\rm BKT}$ is more favorable than as a function of $L/\xi_0$ (see Appendix~\ref{sec:def_cost} for details).

In Figs.~\ref{fig2}(b) and~\ref{fig2}(e) we define the correlation length as $\xi_{\rm BKT} \to$ sign[$W$-$W^*$] $\xi_{\rm BKT}$.
The regime $\xi_{\rm BKT} < 0$ then corresponds to the ergodic regime and $\xi_{\rm BKT} > 0$ to the nonergodic regime.
It is intriguing that by increasing $L$ the results in the scaling solution exhibit a flow towards the ergodic regime, i.e., to the negative values of $\xi_{\rm BKT}$.
We interpret this feature as robustness of many-body quantum chaos.
On a quantitative level, it is manifested as a linear drift of the crossing point $W^*$ with $L$ for the available system sizes, as shown in the inset of Fig.~\ref{fig2}(b).
Note that the dependence $W^*(L)$, described by the symbols in the inset of Fig.~\ref{fig2}(b), is obtained without imposing any functional form of $W^*(L)$, while the solid lines are obtained using the functional form $W^*(L) = w_0 + w_1 L$, with $w_0$ and $w_1$ as fitting parameters.
Both approaches provide a nearly identical proportionality coefficient of the linear drift $W^*(L) \propto L$.

Intriguingly, the crossing points $W^*(L)$ in the inset of Fig.~\ref{fig2}(b) agree both qualitatively and quantitatively with those obtained using other ergodicity indicators such as the eigenstate entanglement entropy $S$ and the average level spacing ratio $r$~\cite{suntajs_bonca_20}, where a much larger interval of disorders has been taken into account.
This suggests that the ergodicity breaking transition described here corresponds to the onset of departure of $S$ from the maximal value (calculated in~\cite{page_93, vidmar_rigol_17}), and of $r$ from the corresponding random matrix theory prediction $r_{\rm GOE}$~\cite{atas_bogomolny_13}, see Fig.~\ref{fig1}.

We complement our study of scaling solutions with the analysis in Figs.~\ref{fig2}(c) and~\ref{fig2}(f), where the ergodicity indicator $g(W,L)$ is plotted as a function of $W/L$.
We observe clear features of a transition for both models, characterized by a constant $g$ at a constant $W^*/L $.
This implies a linear dependence of the crossing point $W^*$ on $L$, which is consistent with the results from the scaling solution of $g(L/\xi_{\rm BKT})$ discussed before.
A similar analysis to those in Figs.~\ref{fig2}(c) and~\ref{fig2}(f) has been recently performed for the three-dimensional (3d) and five-dimensional (5d) Anderson models~\cite{sierant_delande_20}.
It was shown that the critical disorder $W_{\rm c}$ of the Anderson localization transition can be reliably detected using the same methodology as applied here, i.e., by requiring $g = {\rm const}$ when $L$ increases (cf.~the insets of Fig.~1 in~\cite{sierant_delande_20}).
However, in sharp contrast to the results in Figs.~\ref{fig2}(c) and~\ref{fig2}(f), the values of $W_{\rm c}$ at the 3d and 5d Anderson transitions do not exhibit any pronounced system size dependence.

To summarize our results reported so far, there are two important lessons to be learned from Fig.~\ref{fig2}.
First, the ergodicity breaking transition in disordered many-body systems carries analogies with the Anderson localization transition~\cite{sierant_delande_20}, for which the transition point can be detected requiring $\lim_{L\to\infty} g(W^*,L) = g^* =  {\rm const}$.
Second, the scaling solution of $g$ (and of other ergodicity indicators such as $S$ and $r$~\cite{suntajs_bonca_20}) obtained by requiring the optimal data collapse as a function of $L/\xi_{\rm BKT}$, provides an efficient method to pin the ergodicity breaking transition in finite systems.

\begin{figure}[!b]
\begin{center}
\includegraphics[width=0.99\columnwidth]{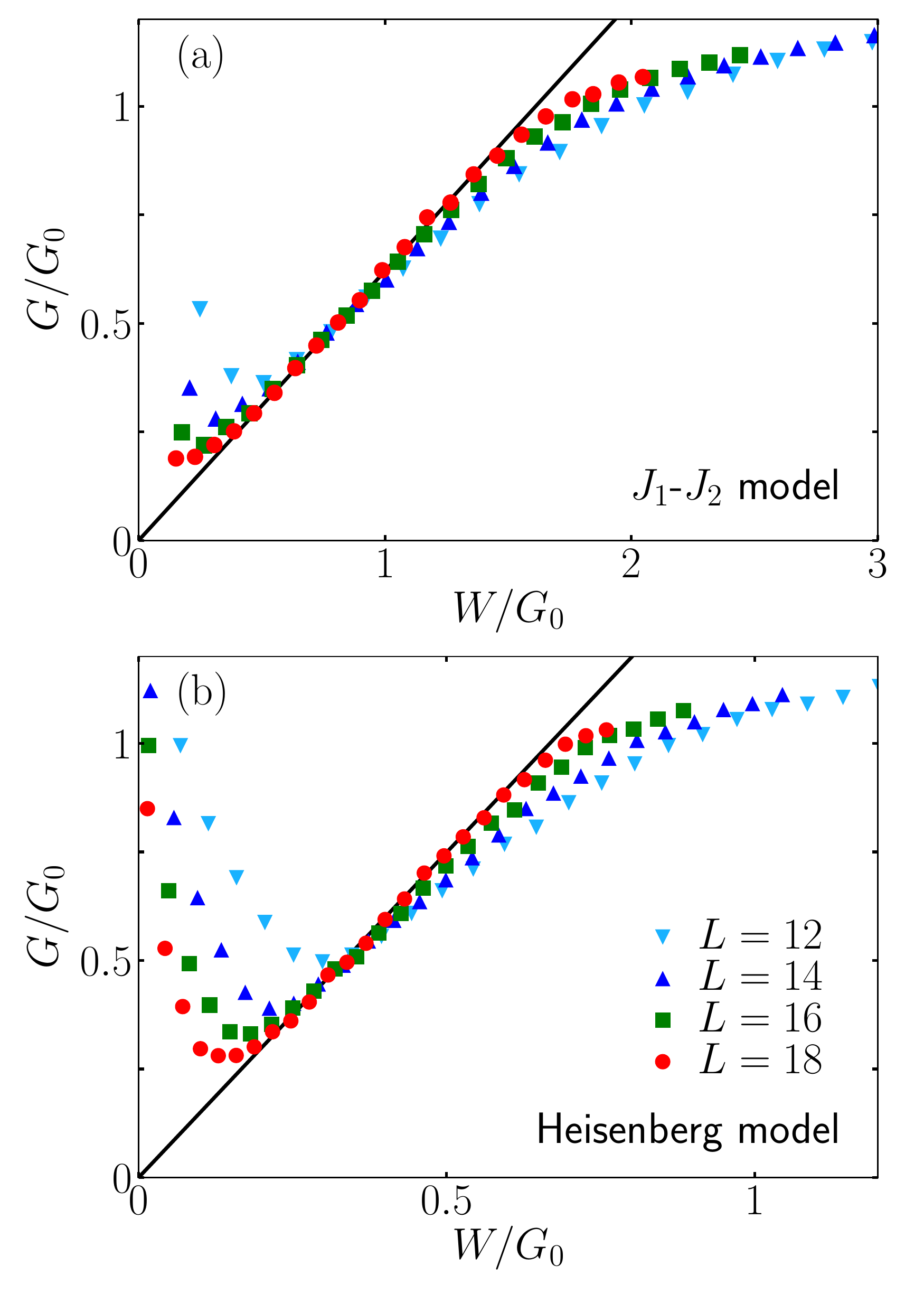}
\end{center}
\caption{
{\it Thouless time $t_{\rm Th}$ in disordered spin chains.}
(a) $J_1$-$J_2$ model and (b) Heisenberg model.
We plot $G/G_0$ as a function of $W/G_0$, where $G = \log_{10}[t_{\rm Th}/(t_0 L^2)]$ and $G_0 = \log_{10}[t_{\rm H}/(t_0 L^2)]$.
The constant $t_0$ is obtained by fitting $t_{\rm Th}$ to Eq.~(\ref{tTh_scaling}) at $L=18$.
We get (a) $t_0 = 1.09 \cdot 10^{-2}/|J_1|$ and $\Omega = 0.70$ [fits for $1.5 \leq W \leq 4$], and (b) $t_0 = 0.87 \cdot 10^{-2}/|J_1|$ and $\Omega = 0.29$ [fits for $1 < W < 2$].
Lines are linear functions $\log_{10}(e)/\Omega \, \times (W/G_0)$.
\label{fig3} }
\end{figure}

\begin{figure*}
\begin{center}
\includegraphics[width=2.08\columnwidth]{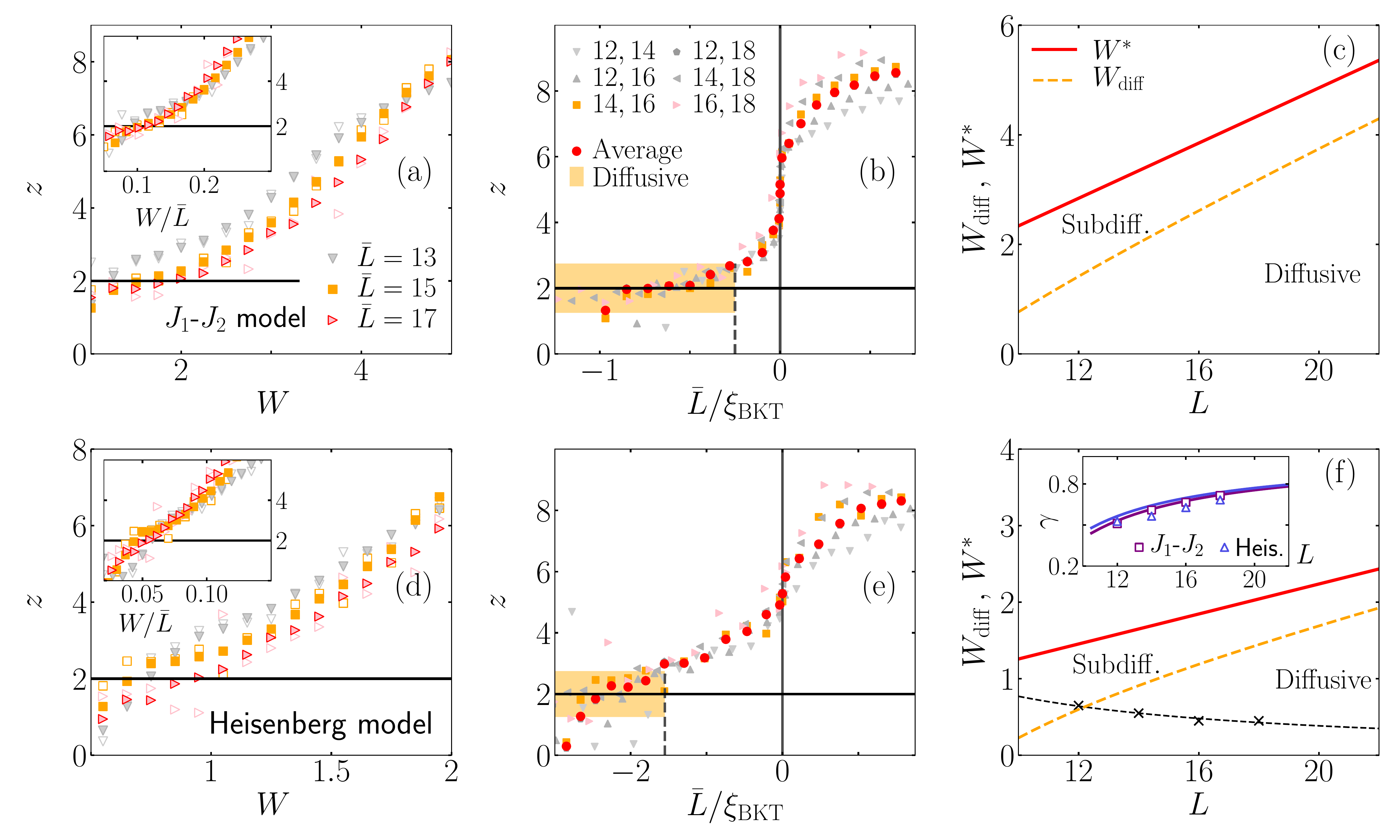}
\end{center}
\caption{
{\it Scaling properties of the Thouless time $t_{\rm Th}$ in disordered spin chains.}
Results are shown for the $J_1$-$J_2$ model (upper panels) and for the Heisenberg model (lower panels).
(a) and (d): The exponent $z(W)$ of $t_{\rm Th}$, as defined in Eq.~(\ref{def_tTh_z}).
The insets show $z(W/\bar L)$.
We calculate $z$ from Eq.~(\ref{zW_extract}) for two consecutive system sizes $L_1$ and $L_2$, with $\bar L = (L_1+L_2)/2$.
Open symbols denote the raw results while filled symbols denote the averages of $z$ over 5 consecutive values of $W$ around the target $W$.
(b) and (e): $z$ as a function of $\bar L/\xi_{\rm BKT}$ for different pairs of system sizes $L_1$ and $L_2$ (indicated in the legend), using identical parameters for $\xi_{\rm BKT}$ as in Fig.~\ref{fig2}.
Red circles denote the average over all pairs of $L_1$ and $L_2$.
Shaded regions mark the regimes of nearly diffusive exponents $z \approx 2$.
(c) and (f): The crossing point $W^*$ [full red line; results from Fig.~\ref{fig2} using the fitting function $W^* = w_0 + w_1 L$] and the diffusion breakdown point $W_{\rm diff}$ [dashed orange line], versus $L$.
$W_{\rm diff}$ is obtained by solving the equation $\bar L/\xi_{\rm BKT} = \rho$, where $\rho$ is marked by vertical dashed lines in (b) and (e), and the parameters in $\xi_{\rm BKT}$ are obtained from the scaling solution in Fig.~\ref{fig2} (see Appendix~\ref{sec:def_cost} for exact values).
We set $\rho = -0.25$ in (b) and $\rho = -1.55$ in (e), which correspond to $g_{\rm diff} \approx 1$ in Figs.~\ref{fig2}(b) and~\ref{fig2}(e).
Stars in (f) corresponds to positions of peaks in $g(L)$ in Fig.~\ref{fig2}(d), while the black dashed line is a guide to the eye.
Symbols in the inset in (f) show the exponent $\gamma$ from Eq.~(\ref{def_eta}) extracted from $W_{\rm diff}(L)$ in (c) and (f), while the lines (the upper line for the Heisenberg model and the lower line for the $J_1$-$J_2$ model) represent $\gamma = 1 - g_{\rm diff}/\log_{10} (t_{\rm H} |J_1|)$ [with $g_{\rm diff} = 1$], yielding good agreement.
Note that the latter expression suggests the large $L$ dependence $\gamma \approx 1 - {\rm const}/L$.
\label{fig4} }
\end{figure*}

{\bf Scaling of Thouless time below the transition.}
Finally, we address the question about the impact of disorder on the ergodic side of the transition.
The main question is to what extent does the Thouless time $t_{\rm Th}$ exhibit a diffusive character, and whether one can detect fingerprints of a subdiffusive transport.
The latter was argued to be a characteristic feature of moderately disordered systems~\cite{barlev_cohen_15, agarwal_gopalakrishnan_15, luitz_laflorencie_16, khait_gazit_16, znidaric_scardicchio_16, luitz_barlev_17}.

We start our analysis by exploring the diffusive scaling of $t_{\rm Th}$, described by $t_{\rm Th} \propto L^2$.
To this end, we calculate $G = \log_{10}[t_{\rm Th}/(t_0 L^2)]$, where $t_0$ is a characteristic time in units of $\hbar/|J_1|$ (we set $\hbar \equiv 1$ further on), and normalize it by the corresponding value for the Heisenberg time $G_0 = \log_{10}[t_{\rm H}/(t_0 L^2)]$.
Results shown in Figs.~\ref{fig3}(a) and~\ref{fig3}(b) exhibit a good collapse of $G/G_0$ versus $W/G_0$ for different system sizes at weak disorder (away from the $W \to 0$ limit).
In addition, we observe a linear increase of $G/G_0$ with $W/G_0$, which suggests an exponential growth of $t_{\rm Th}$ with disorder $W$.
We express these two observations by the ansatz for $t_{\rm Th}$ in which the $L$ and $W$ dependences are decoupled,
\begin{equation} \label{tTh_scaling}
 t_{\rm Th} = t_0 \, e^{W/\Omega} \, L^{2} \, ,
\end{equation}
and $\Omega$ is a constant.
While the results in Figs.~\ref{fig3}(a) and~\ref{fig3}(b) show a progressive dispersion of the data collapse with increasing $W$,
we also highlight that the results for different $L$ seem to exhibit a linear increase up to similar values of $W/G_0$.
Note that the latter scales with the system size as $W/G_0 \propto W/L$.
The condition $W/G_0 = {\rm const}$ therefore gives rise to a similar scaling with the system size as observed for the crossing point $W^*$ in the inset of Fig.~\ref{fig2}(b).

Existence of the ergodic regime with diffusive transport and the exponential decay of the inverse Thouless time $t_{\rm Th}^{-1}$ with disorder $W$ may be consistent with some of previous works on the diffusion constant and dc conductivity, which also decreases exponentially with $W$~\cite{barisic_prelovsek_10, steinigeweg_herbrych_16, barisic_kokalj_16, prelovsek_mierzejewski_17}.
However, for $W \gg \Omega$ the inverse Thouless time in finite systems may become comparable to the mean level spacing and therefore difficult to be detected from spectral analysis.
Moreover, even if the spectral form factor suggests the diffusive scaling of $t_{\rm Th}$, this may not exclude an emergence of subdiffusion at short times, detected by other probes.
Indeed, several previous works~\cite{barlev_cohen_15, khait_gazit_16, znidaric_scardicchio_16} highlighted an emergence of subdiffusion at rather weak disorders (or eventually even at $W \to 0$).
We study the latter scenario by testing a more general ansatz for $t_{\rm Th}$,
\begin{equation} \label{def_tTh_z}
t_{\rm Th}=c(W) \, L^{\,z(W)} \,,
\end{equation}
where the exponent $z$ is reminiscent of the dynamical exponent whose disorder dependence was subject of intense research~\cite{luitz_barlev_17}.
The scaling in Eq.~(\ref{def_tTh_z}) is consistent with subdiffusion for $z > 2$.
We numerically extract $z(W)$ from pairs of system sizes $L_1$ and $L_2$ as
\begin{equation} \label{zW_extract}
 z(W) = \frac{\log_{10}[t_{\rm Th}(L_2)/t_{\rm Th}(L_1)]}{\log_{10}[L_2/L_1]} \,,
\end{equation}
where the mean system size is $\bar L = (L_1 + L_2)/2$.
Results shown in Figs.~\ref{fig4}(a) and~\ref{fig4}(d) suggest two observations: the emergence of the regime with $z>2$ at large enough disorder, and a drift of $z(W)$ with $\bar L$.
The drift may be eliminated when plotting $z(W/\bar L)$, see the insets of Figs.~\ref{fig4}(a) and~\ref{fig4}(d).
However, we also argue that it may be difficult to predict the optimal data collapse due to the limitation to a rather narrow interval of system sizes $\bar L = 13,15,17$.

Results reported so far raise the question about the boundaries of the disorder regime in which the scaling properties of $t_{\rm Th}$ may be consistent with diffusion. 
We denote $W_{\rm diff}$ as the disorder value for which the breakdown of the diffusive scaling is observed upon increasing the disorder.
In general, extracting the scaling $W_{\rm diff}(L)$ is a daunting task that is rarely addressed in the literature (with some exceptions~\cite{bera_detomasi_17}).
Here we make an approximation inspired by the scaling solution of the ergodicity indicator $g$ in Fig.~\ref{fig2}.
We plot $z$ as a function of $\bar L/\xi_{\rm BKT}$, and make the link between the breakdown of the diffusive scaling and the values of $\bar L/\xi_{\rm BKT}$.
A rather good collapse of results in Figs.~\ref{fig4}(b) and~\ref{fig4}(e) for different $\bar L$ suggest that this approximation is reasonable.
Then, we plot the values of $W_{\rm diff}(L)$ along with the crossing point function $W^*(L)$ in Figs.~\ref{fig4}(c) and~\ref{fig4}(f).
Results show (consistently with the previous work~\cite{luitz_barlev_17}) that  the regime of diffusive scaling is limited to relatively small values of $W$, in particular in the Heisenberg model (in the latter, we also observe the tendency $z < 2$ at small disorder due to proximity to the integrable point).
At the same time, $W_{\rm diff}$ increases (roughly linearly) with $L$ for system sizes under investigation.
This is a consequence of the observation in Figs.~\ref{fig4}(b) and~\ref{fig4}(e) [see the vertical dashed lines] that the breakdown of the diffusive scaling occurs roughly at $\bar L/\xi_{\rm BKT} = {\rm const}$, where $\xi_{\rm BKT}$ contains the linear drift of the crossing point $W^*(L)$.

Perhaps even more instructive is to identify the value of $t_{\rm Th}$ at $W_{\rm diff}$, i.e., at the point when the diffusive scaling breaks down.
We make a power-law ansatz $t_{\rm Th}(W_{\rm diff}) |J_1| = \left(t_{\rm H}(W_{\rm diff}) |J_1|\right)^\gamma$ and calculate the power $\gamma$,
\begin{equation} \label{def_eta}
 \gamma = \frac{\log_{10} [t_{\rm Th}(W_{\rm diff})|J_1|]}{\log_{10} [t_{\rm H}(W_{\rm diff}) |J_1|]} \,,
\end{equation}
shown in the inset of Fig.~\ref{fig4}(f).
Results suggest that for system sizes available from exact diagonalization ($10 \lesssim L \lesssim 20$), the breakdown of the diffusive scaling occurs at $0.5 \lesssim \gamma \lesssim 0.8$, where $\gamma$ increases with $L$.
We highlight that rather than studying the existence of diffusion in different regimes of model parameters, it may be more instructive to explore the existence of diffusion in finite systems as a function of the ratio $t_{\rm Th}/t_{\rm H}$.
The latter approach may be general and more easily contrasted to other ergodicity breaking scenarios such as approaching the Bethe ansatz integrable point.

{\bf Discussion.}
Our work studies the breakdown of ergodicity in disordered quantum many-body systems from the perspective of spectral properties beyond the level spacing analysis.
We argue that the ergodicity breaking transition occurs when values of the Thouless time and the Heisenberg time become comparable, $t_{\rm Th} \approx t_{\rm H}$, and hence the ergodicity indicator $g$ becomes a system-size independent constant that is equal or close to zero.
Using an identical criterion one can reliably detect the well established Anderson localization transition in three (and higher) dimensions~\cite{sierant_delande_20}, thereby suggesting a link between the ergodicity breaking transition in disordered many-body systems and the Anderson localization transition.

In the context of interacting spin-1/2 chains in the total $s^z=0$ magnetization sector, we report the scaling solution of the ergodicity indicator $g$ across the disorder induced ergodicity breaking transition, which allows for a detection of the crossing point $W^*$ in finite systems.
The crossing point is located at disorder values much lower than those reported in previous studies using comparable system sizes~\cite{oganesyan_huse_07, pal_huse_10, barkelbach_reichman_10, Luitz2015, bertrand_garciagarcia_16, khemani_lim_17, gray_bose_18, mace_alet_19}.
However, for system sizes under investigation, the scaling solution exhibits robustness of the ergodic regime since the crossing point $W^*$ increases linearly with the system size $L$.

One of the most widely studied disordered models so far is the isotropic Heisenberg chain, for which our results predict the ergodicity breaking transition to occur at $W^* \approx 2$ for system sizes $L \lesssim 20$.
A quantitative comparison with other studies of the same model (for similar system sizes) suggests the following features to occur at roughly the same disorder:
the eigenstate entanglement entropy and the level spacing ratio start deviating from the random-matrix-theory predictions~\cite{suntajs_bonca_20},
a gap opens in the one-particle density-matrix spectrum of eigenstates~\cite{bera_schomerus_15}, 
the Schmidt gap opens in the spectrum of reduced density matrices of eigenstates~\cite{gray_bose_18},
correlations between nearby eigenstates (described by the Kullback-Leibler divergence) change quantitatively~\cite{Luitz2015},
and properties of the average and typical offdiagonal matrix elements of local observables depart from each other~\cite{serbyn_papic_17}.
This suggests a rich variety of phenomena occurring at the ergodicity breaking transition invoked by the criterion $t_{\rm Th} \approx t_{\rm H}$.

Another intriguing question is the nature of states on the nonergodic side of the ergodicity breaking transition, i.e., for the disorder values $W > W^*$.
A recent study based on a similar scaling solution as applied here~\cite{suntajs_bonca_20} argued that the eigenstate entanglement entropy increases as a volume law (with a subthermal prefactor) even for $W > W^*$.
A possible scenario is then that the MBL transition (i.e., a transition to the regime of area-law entanglement) occurs at even higher disorders not studied here, for which $t_{\rm Th} \gg t_{\rm H}$.
We note, however, that numerical exact diagonalization methods provided remarkable benchmarks in the past (with negligible system-size effects) to unveil fingerprints of eigenstate thermalization and quantum chaos in the opposite limit $t_{\rm Th} \ll t_{\rm H}$ (see, e.g.,~\cite{dalessio_kafri_16, mondaini_rigol_17, jansen_stolpp_19} for examples in disorder-free many-body systems).
A challenge for future studies is therefore to better understand the robustness of exact numerical approaches for many-body lattice systems in the regime $t_{\rm Th} \gg t_{\rm H}$.

Finally, we also explore the signatures of diffusion in the scaling of the Thouless time $t_{\rm Th}$.
We show that the breakdown of the diffusive scaling occurs in finite systems at disorder values that are smaller than those at the ergodicity breaking transition.
However, the breakdown of the diffusive scaling is also associated with pronounced system-size effects.
This may be consistent with the emergence of subdiffusion at much larger disorder values, or with scenarios~\cite{bera_detomasi_17} in which subdiffusion is a transient phenomenon.
An outstanding goal for future studies is to establish quantitative links between the inverse Thouless time (the Thouless energy), and the diffusion constant (or the dc optical conductivity~\cite{barisic_prelovsek_10, steinigeweg_herbrych_16, barisic_kokalj_16, prelovsek_mierzejewski_17}), which has for many-body systems remained elusive.

Our work studies the impact of random on-site disorder on the ergodic spectral properties of certain interacting spin-1/2 chains in the total $s^z=0$ magnetization sector.
In future work, it would be exciting to carry out analogous analyses in similar systems such as the quasidisordered systems, the disordered Floquet systems, and the interacting spin-1/2 systems where the only conserved quantity is the total energy.
This would provide a more comprehensive view on the robustness of the many-body quantum chaos and the thermodynamic (in)stability of the many-body localization in different families of the disordered many-body quantum systems.

\begin{acknowledgments}
We acknowledge insightful discussions with W. De Roeck, F. Heidrich-Meisner, D. Luitz, A. Polkovnikov, P. Prelov\v sek, M. Rigol, T. H. Seligman, D. Sels and M. \v Znidari\v c.
This work was supported by the Slovenian Research Agency (ARRS), Research Core Fundings Grants No.~P1-0044, No.~P1-0402 and No.~J1-1696, and by European Research Council (ERC) under Advanced Grant 694544 -- OMNES.
\end{acknowledgments}

\appendix

\section{Calculation of the Heisenberg time} \label{sec:def_tH}

The Heisenberg time $t_{\rm H}$ is defined as the inverse mean level spacing, $t_{\rm H} = \hbar/\overline{\delta E}$, where $\hbar \equiv 1$.
We define the mean level spacing $\overline{\delta E} = \Gamma_0/(\chi {\cal D})$, where
\begin{equation} \label{def_gamma}
\Gamma_0^2 = \langle {\rm Tr} \{ \hat H^2 \}\rangle/{\cal D} - \langle {\rm Tr} \{ \hat H \}^2 \rangle /{\cal D}^2
\end{equation}
is the variance after disorder averaging, ${\cal D} = \binom{L}{L/2}$ is the Hilbert space dimension, and $\chi$ controls the number of energy levels in the interval $[\bar E, \bar E + \Gamma_0]$, with $\bar E = \langle {\rm Tr} \{ \hat H \} \rangle /{\cal D}$.
We calculate $\chi$ using the Gaussian approximation for the density of states $\chi = \int_0^{\Gamma_0} \exp\{-E^2/(2\Gamma_0^2) \} /(\sqrt{2\pi}\Gamma_0) dE$, which yields $\chi = (1/2)\mbox{erf}[1/\sqrt{2}] \approx 0.3413$.
In Eq.~(\ref{def_gamma}) we denote $\langle \cdots \rangle$ as the averaging over disorder realizations.

We numerically calculate $\Gamma_0$ in Eq.~(\ref{def_gamma}) for given Hamiltonians in the $s^z = 0$ magnetization sector.
We verified that those values of $\Gamma_0$ are very close to the ones in the grandcanonical ensemble, where
$\Gamma_0^2 = L[(J_1^2 + J_2^2)/8 + ((J_1\Delta)^2 + (J_2 \Delta)^2)/16 + (J_1 W)^2/12]$ (relative differences are of the order $10^{-2}$).
We use the latter $\Gamma_0$ to obtain the continuous curve $\gamma(L) = 1 - 1/\log_{10}(t_{\rm H} |J_1|)$ in the inset of Fig.~\ref{fig4}(f).

\section{Calculation of the Thouless time from the Spectral form factor} \label{sec:def_tTh}

We extract the Thouless time $t_{\rm Th}$ from the spectral form factor (SFF).
The latter is a Fourier transform of the spectral two-point correlations, defined as
\begin{equation} \label{def_Kt}
\begin{split}
K(\tau) &= \frac{1}{Z} \left\langle \left|\sum_{\alpha = 1}^{\cal D} \rho(\varepsilon_\alpha) e^{-i 2\pi\varepsilon_\alpha \tau}\right|^2 \right\rangle \, ,
\end{split}
\end{equation}
where $\{ \varepsilon_1\le\varepsilon_2\le \cdots\varepsilon_{\cal D} \}$ is the complete set of Hamiltonian eigenvalues after spectral unfolding and $\tau$ is the scaled time.
We choose the normalization $Z$ is such that $K(\tau\gg 1) \simeq 1$ and we use a smooth filter $\rho(\varepsilon)$ to eliminate contributions of spectral edges (see details below).
The SFF $K(\tau)$ in Eq.~(\ref{def_Kt}) can also be referred to as the unconnected SFF.
A comparison to the connected SFF is carried out in Appendix~\ref{sec:def_connected}.

We eliminate the influence of the local density of states by spectral unfolding.
It transforms an ordered set of Hamiltonian eigenvalues $\{ E_\alpha \}$ to an ordered set of unfolded eigenvalues $\{ \varepsilon_\alpha \}$, for which the local mean level spacing is set to unity at all energy densities.
This implies ${\cal N}^{-1} \sum_\alpha \delta \varepsilon_\alpha =1$, where the average is performed in a microcanonical window, ${\cal N}$ is the number of elements in the window, and $\delta \varepsilon_\alpha = \varepsilon_{\alpha+1}-\varepsilon_\alpha$.
The scaled Heisenberg time (i.e., the inverse mean level spacing of the unfolded spectrum) is therefore $\tau_{\rm H} = 1$.
Following the standard unfolding procedure, we introduce the cumulative spectral function ${\cal G}(E) = \sum_\alpha  \Theta(E-E_\alpha)$, where $\Theta$ is the unit step function.
The stepwise distribution function is then smoothed out by fitting a polynomial $\bar g_n(E)$ of degree $n$ to ${\cal G}(E)$ and the unfolded eigenvalues are defined as $\varepsilon_\alpha = \bar g_n(E_\alpha)$.
We verified that using polynomials $\bar g_n(E)$ of different degrees $n$ in the unfolding procedure does not affect the final results.
We used $n=10$ in calculations of the spectral form factor $K(\tau)$.

The role of the filtering function $\rho(\varepsilon_\alpha)$ in Eq.~(\ref{def_Kt}) is to avoid effects from the spectral edges.
Its specific form should not influence the main features of $K(\tau)$, provided that the filtering function is smooth enough (e.g.~analytic), symmetric with respect to the center of the unfolded spectrum, and has a vanishingly small amplitude at the spectral edges.
In our calculations we first perform spectral unfolding for each disorder realization separately.
Then we filter each unfolded spectrum using a Gaussian filter
$\rho(\varepsilon_\alpha) = \exp\{-\frac{\left(\varepsilon_\alpha - \bar\varepsilon \right)^2}{2\left(\eta\Gamma\right)^2}\}$, where $\bar\varepsilon$ and $\Gamma^2$ are the average energy and the variance, respectively, at a given disorder realization, and $\eta$ a dimensionless parameter that controls the effective fraction of eigenstates included in $K(\tau)$.
We set $\eta=0.5$ in the calculations presented in the main text.
The overwhelming majority of eigenstates in our analysis are hence those that govern the system properties at infinite temperature.
To ensure proper normalization, yielding $K(t\gg 1)\simeq 1$ in general and $K(t)\equiv 1$ for Poisson random spectrum, we then set $Z = \langle \sum_\alpha \left|\rho(\varepsilon_\alpha)\right|^2 \rangle$.

The energy spectra used in the $K(\tau)$ calculations were obtained by means of full numerical exact diagonalization, which was performed for $N_\mathrm{samples}$ different disorder realizations for each value of $W$ and $L$.
We denote $\overline{N_\mathrm{samples}}$ as the number of disorder realizations averaged over all $W$ at a fixed $L$.
We used $\overline{N_\mathrm{samples}} = 1000, 949, 882, 332$ for $L=12, 14,16,18$ in the $J_1$-$J_2$ model,
and $\overline{N_\mathrm{samples}} = 1000, 445, 405, 100$ for $L=12, 14,16,18$ in the Heisenberg model, respectively.

Numerical results for $K(\tau)$ are contrasted to predictions of the random matrix theory, in particular the result for the Gaussian orthogonal ensemble (GOE) for $\tau < 1$,
\begin{equation} \label{def_Kgoe}
K_{\rm GOE}(\tau) = 2\tau - \tau \ln(1+2\tau) \,,
\end{equation}
applicable for systems with time-reversal symmetry~\cite{mehta_91}.
Using $K_{\rm GOE}$, we define the scaled Thouless time $\tau_{\rm Th}$ as the onset time of a universal linear ramp in $K(\tau)$, i.e., we define $\tau_{\rm Th}$ such that for $\tau > \tau_{\rm Th}$, we obtain $K(\tau) \simeq K_{\rm GOE}(\tau) \simeq 2\tau$
(see the open circles in Fig.~\ref{fig1}).
Note that such extraction of $\tau_{\rm Th}$ from $K(\tau)$ in finite systems is not sharply defined due to the noise in $K(\tau)$ when it approaches $K_{\rm GOE}(\tau)$.
In Appendix~\ref{sec:protocol} we describe the detailed protocol (including specific examples) of the Thouless time extraction.

Finally, we calculate the Thouless time in physical units as $t_{\rm Th} = \tau_{\rm Th} t_{\rm H}$.
Note that the ratio between the Thouless and the Heisenberg time (and hence the ergodicity indicator $g$ in Fig.~\ref{fig2}) is identical in both physical and scaled units, $t_{\rm Th}/t_{\rm H} = \tau_{\rm Th}/\tau_{\rm H}$.
However, quantities used in Figs.~\ref{fig3} and~\ref{fig4} require characteristic times in physical units as an input.

\section{Average level spacing ratio} \label{sec:def_r}

In the inset of Fig.~\ref{fig1} we also show the average level spacing ratio $r$.
To obtain $r$, one first defines
\begin{equation} \label{def_rtilde}
  \tilde{r}_\alpha = \frac{\min \{\delta E_\alpha, \delta E_{\alpha-1} \}} {\max \{\delta E_\alpha, \delta E_{\alpha-1}\}} = \min \{ r_\alpha, r_\alpha^{-1} \} \,
\end{equation}
for a target eigenstate $|\alpha\rangle$, where $r_\alpha$ is the ratio of consecutive level spacings, $r_\alpha = \delta E_\alpha/\delta E_{\alpha-1}$, and $\delta E_\alpha = E_{\alpha+1} - E_\alpha$ is the level spacing.
Hence, no unfolding is necessary to eliminate the influence of finite-size effects through the local density of states.
We obtain $r$ by first averaging $\tilde{r}_\alpha$ over $N_\mathrm{ev}=500$ eigenstates near the center of the spectrum for every disorder realization, and then over an ensemble of spectra for different disorder realizations, with $\overline{N_{\rm samples}} = 980$ at $L=18$.
The GOE prediction for $r$ is $r_{\rm GOE} \approx 0.5307$~\cite{atas_bogomolny_13}, while the prediction for energy levels with Poisson statistics is $r_{\rm Poisson} \approx 0.3863$~\cite{oganesyan_huse_07}.

\section{Cost function minimization approach} \label{sec:def_cost}

Figures~\ref{fig2}(b) and~\ref{fig2}(e) show the optimal scaling collapses of $g(W,L)$ for the $J_1$-$J_2$ and the Heisenberg model, respectively, as functions of $L/\xi_{\rm BKT}$.
The algorithm to obtain the optimal data collapse is explained below.

In general, we consider as the correlation length $\xi$ either the BKT correlation length $\xi_{\rm BKT}$ from Eq.~(\ref{def_xi_kt}) or the power-law correlation length $\xi_0 = |W-W^*|^{-\nu}$.
Then, we seek for the scaling collapses as functions of $L/\xi$ using the optimal fitting parameters of $\xi$.
The fitting parameters are:
the coefficient $b$ in $\xi_{\rm BKT}$ ($\nu$ in $\xi_0$), and the parameters of the crossing point fitting function $W^*(L)$.
For the latter we consider two cases:
(i) the most general case (free of any functional form) in which independent fitting parameters $w^*(L)$ are used, i.e., four different parameters for four different lattice sizes $L=12,14,16,18$;
and (ii) the crossing point fitting function $W^*(L) = w_0 + w_1 L$, with free parameters $w_0$ and $w_1$.

Following Ref.~\cite{suntajs_bonca_20}, we obtain the quality of the data collapse using the cost function ${\cal C}_g$
\begin{equation} \label{def_costfun}
 {\cal C}_g = \frac{ \sum_{j=1}^{N_{\rm p}-1} |g_{j+1} - g_{j}| }{\max\{g_j\} - \min\{g_j\}} - 1 \,,
\end{equation}
where $g_j$ represent values of $g$ at different $W$ and $L$.
Specifically, we sort all $N_{\rm p}$ values of $g_j$ according to nondecreasing values of ${\rm sign}[W$-$W^*] L/\xi$.
Then, using a differential evolution method, we obtain the best data collapse by finding the parameters in $W^*(L)$ and $\xi(W,L)$ for which the cost function ${\cal C}_g$ is minimal.
In the procedure, we include results of $g$ within the interval $1 < W < 6$ for the $J_1$-$J_2$ model and within the interval $0.6 < W < 2.5$ for the Heisenberg model.
This results in $N_{\rm p} = 76$ points included in the cost function minimization procedure for both the $J_1$-$J_2$ and the Heisenberg model.

For both models under investigation, we consistently find that:
(i) the scaling solution using $L/\xi_{\rm BKT}$ is better than the solution using $L/\xi_0$.
In particular, using independent fitting parameters $w^*(L)$ we get ${\cal C}_g(\xi_{\rm BKT}) = 0.036$ and ${\cal C}_g(\xi_0) = 0.051$ for the $J_1$-$J_2$ model and 
${\cal C}_g(\xi_{\rm BKT}) = 0.087$ and ${\cal C}_g(\xi_0) = 0.126$ for the Heisenberg model.
(ii) The independent fitting parameters $w^*(L)$ (using both $\xi_{\rm BKT}$ and $\xi_0$) are very well described by a function that increases linearly with $L$.
As shown in the inset of Fig.~\ref{fig2}(b), the slope of the linear increase with $L$ of $w^*(L)$ is nearly identical to $w_1$ of the fitting function $W^* = w_0 + w_1 L$.
In the latter case, using $\xi_{\rm BKT}$ we get from the cost function minimization $w_0 = -0.181$, $w_1 = 0.252$, $b=4.62$ for the $J_1$-$J_2$ model and $w_0 =0.278 $, $w_1 = 0.098$, $b=1.89$ for the Heisenberg model.
The corresponding values of $w_1$ are also used to plot the vertical lines in Figs.~\ref{fig2}(c) and~\ref{fig2}(f).

We note that the overwhelming majority of previous numerical analyses  of the ergodicity indicators across the ergodicity breaking transitions considered scaling solutions as a function of $\chi L^{1/\nu}$, where the linear scaling variable is $\chi \propto \xi_0^{-1/\nu} = |W - W^*|$.
Here we extend those analyses by also considering a specific nonlinear scaling variable $\chi \propto \xi_{\rm BKT}^{-1} = e^{-b/\sqrt{|W-W^*|}}$ and study the scaling solution as a function of $L/\xi_{\rm BKT}$.
Furthermore, by introducing an additional $L$ dependence of the crossing point $W^*(L) = w_0 + w_1 L$ we obtain excellent scaling collapses of the ergodicity indicator $g$ as a function of $L/\xi_{\rm BKT}$, as shown in Figs.~\ref{fig2}(b) and~\ref{fig2}(e) for both models under investigation.
These scaling collapses are obtained using only three free fitting parameters $w_0$, $w_1$ and $b$, as explained in the paragraph above.
This number is much smaller than the typical number of free fitting parameters invoked in the scaling analyses where the additional $L$ dependence is introduced through an irrelevant scaling variable (see, e.g., Ref.~\cite{slevin_ohtsuki_99} as an example for the Anderson localization transition).

\section{Connected versus unconnected spectral form factor (SFF)} \label{sec:def_connected}

The results of our calculations of the SFF $K(\tau)$, shown in Fig.~\ref{fig1}, were calculated according to the unconnected SFF $K(\tau)$ defined in Eq.~\eqref{def_Kt}.
Here we also compare the SFF $K(\tau)$ with the \emph{connected} SFF, $K_\mathrm{c}(\tau).$ The latter is obtained by subtracting a nonuniversal disconnected part from $K(\tau)$,
\begin{align} \label{def_Ktc}
K_\mathrm{c}(\tau) = &  \frac{1}{Z}\bigg(\left\langle\left|\sum_{\alpha = 1}^{\cal D} \rho(\varepsilon_\alpha) e^{-i2\pi\varepsilon_\alpha \tau}\right|^2 \right\rangle - \nonumber \\
 & \frac{A}{B}\left|\left\langle \sum\limits_{\alpha=1}^{\cal{D}} \rho(\varepsilon_\alpha)e^{-i2\pi\varepsilon_\alpha \tau} \right\rangle \right|^2\bigg) \, ,
\end{align}
where $A$ and $B$ are the normalization constants ensuring the vanishing of $K_\mathrm{c}(\tau)$ in the $\tau \to 0$ limit, accounting for the spectral filtering
\begin{equation}
A = \left\langle\left|\sum_{\alpha = 1}^{\cal D} \rho(\varepsilon_\alpha)\right|^2 \right\rangle \, , \hspace{5mm}
B = \left|\left\langle \sum\limits_{\alpha=1}^{\cal{D}} \rho(\varepsilon_\alpha) \right\rangle \right|^2 \, .
\end{equation}

\begin{figure}[b]
\begin{center}
\includegraphics[width=1.00\columnwidth]{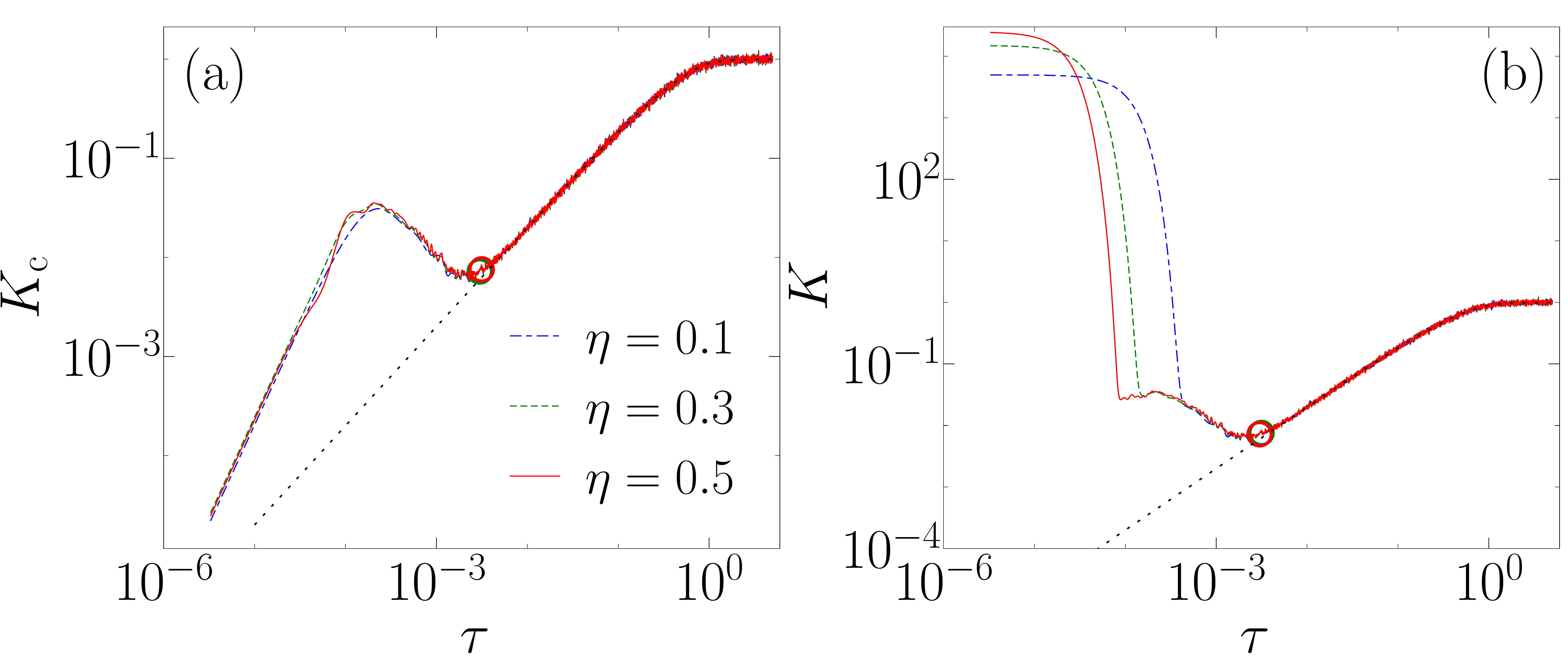}
\end{center}
\caption{{\it Comparison between the connected and unconnected SFF, using different spectral filters.}
(a) The connected SFF $K_{\rm c}(\tau)$, defined in Eq.~(\ref{def_Ktc}).
(b) The SFF $K(\tau)$, defined in Eq.~\eqref{def_Kt}.
We use a Gaussian spectral filter $\rho(\varepsilon)$ introduced in Eq.~\eqref{def_Kt}, where $\eta$ determines the effective width of the filter.
Here we compare results for $\eta=0.1, 0.3$ and 0.5.
Open circles denote the scaled Thouless time $\tau_{\rm Th}$ obtained by using the criterion $\log_{10}[K(\tau_{\rm Th})/K_{\rm GOE}(\tau_{\rm Th})] = 0.08$.
The dotted black line is the GOE prediction $K_{\rm GOE}(\tau)$.
Results are shown for the $J_1$-$J_2$ model at $L=18$ and $W=1$.
\label{figS1} }
\end{figure}

Using as an example the $J_1$-$J_2$ model at $L=18$ and $W=1$, we show in Fig.~\ref{figS1} the difference between $K_{\rm c}(\tau)$ [Fig.~\ref{figS1}(a)] and $K(\tau)$ [Fig.~\ref{figS1}(b)].
While the short time nonuniversal behavior of $K_{\rm c}(\tau)$ and $K(\tau)$ is clearly different, they both follow the GOE prediction $K_{\rm GOE}(\tau)$ after the scaled Thouless time $\tau_{\rm Th}$, which is almost identical for $K_{\rm c}(\tau)$ and $K(\tau)$.
In Fig.~\ref{figS1}, we also show results for different widths $\eta$ of a Gaussian spectral filter $\rho(\varepsilon)$ introduced in Eq.~(\ref{def_Kt}).
Results for $\tau_{\rm Th}$ appear to be fairly independent of the width $\eta$ of the Gaussian filter used in our calculations.
Note, however, that $\eta \lesssim 0.5$ is needed to eliminate contributions from eigenstates at the spectral edges.
Since, in general, one wishes to include as many eigenstates as possible in the analysis, we find that setting $\eta=0.5$ is the optimal parameter choice of the Gaussian spectral filtering.
Further comparison between the results for $\tau_{\rm Th}$ using $K(\tau)$ and $K_{\rm c}(\tau)$, as well as different widths of the spectral filter, is provided in Fig.~\ref{figS3} of Appendix~\ref{sec:robustness}.

\begin{figure*}
\begin{center}
\includegraphics[width=2.00\columnwidth]{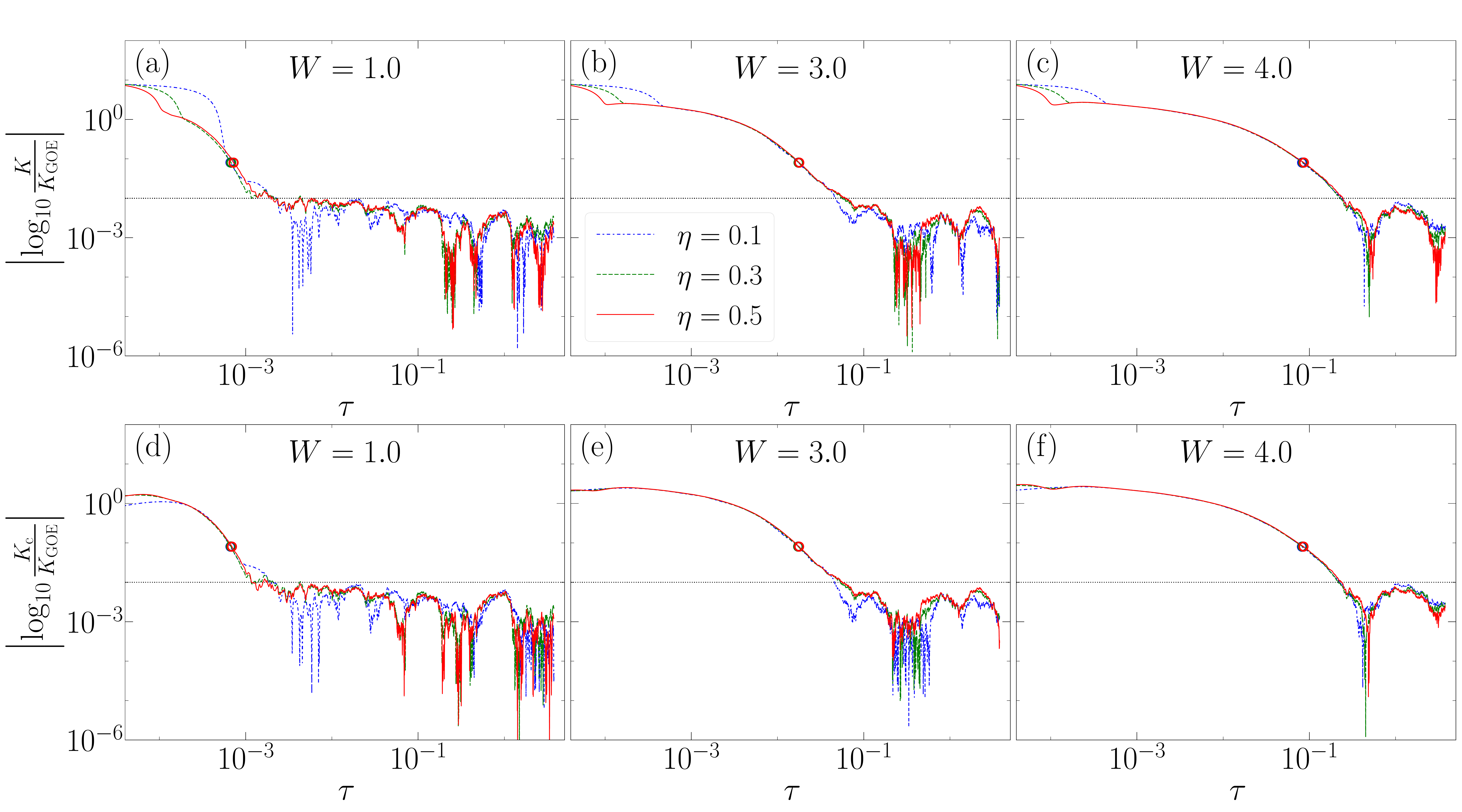}
\end{center}
\caption{{\it Extraction of the Thouless time.}
In the upper row [(a)-(c)] we plot the deviation measure $\Delta K(\tau)$, see Eq.~(\ref{def_deviation}), to quantify the difference between the SFF $K(\tau)$ and the GOE prediction $K_{\rm GOE}(\tau)$.
The lower row [(d)-(f)] shows the analogous quantity for the connected SFF $K_{\rm c}(\tau)$.
Results are shown for the $J_1$-$J_2$ model with the disorder strengths $W=1$ [(a), (d)], $W=3$ [(b), (e)] and $W=4$ [(c), (f)].
In all panels we show results for three different widths $\eta = 0.1, 0.3, 0.5$ of the Gaussian filter $\rho(\varepsilon)$.
Circles denote the scaled Thouless time $\tau_{\rm Th}$ and the horizontal lines denote $\Delta K = 10^{-2}$.
\label{figS2} }
\end{figure*}

\section{Protocol to extract the scaled Thouless times from the SFF} \label{sec:protocol}

Here we describe the protocol to extract the scaled Thouless time $\tau_{\rm Th}$ from the SFF $K(\tau)$.
[An analogous procedure is applied for the connected SFF $K_{\rm c}(\tau)$].
The main goal of the protocol is to obtain robust results for the Thouless time for both $K(\tau)$ and $K_{\rm c}(\tau)$, and different widths $\eta$ of the Gaussian spectral filter.
This is achieved using identical parameters of the protocol for all system sizes $L$, disorders $W$, and for both models under investigation.

We apply the following steps of the protocol: \\
(i) Each $K(\tau)$ curve is calculated for 5000 times $\tau_i$ in the window $\tau_i \in [1/(2\pi {\cal D}),5]$, with $\tau_i$ being equidistant in the logarithmic scale.
We then smoothen out random fluctuations in $K(\tau)$ by calculating a running mean such that each new $K(\tau_i)$ is the average over 200 nearest values of $K(\tau_i)$, and hence the final number of data points is reduced to 4801. \\
(ii) We analyze the difference between $K(\tau)$ and the GOE prediction $K_{\rm GOE}(\tau) = 2\tau - \tau \ln(1+2\tau)$ using the deviation measure
\begin{equation} \label{def_deviation}
\Delta K(\tau) = \left|\log_{10}\frac{K(\tau)}{K_\mathrm{GOE}(\tau)} \right|\;; \;\;\;\;\;\;\;\;\; \Delta K(\tau_{\rm Th}) = \epsilon \, .
\end{equation}
As advanced in Eq.~(\ref{def_deviation}), we then define the scaled Thouless time $\tau_{\rm Th}$ as the time at which $\Delta K(\tau)$ becomes larger than some threshold value $\epsilon$ upon decreasing $\tau$ from the regime $\tau > 1$.
In general, the agreement of $K(\tau)$ with $K_{\rm GOE}(\tau)$ is in finite systems associated with the emergence of noisy data in $\Delta K(\tau)$, with $\Delta K(\tau) \ll 1$.
The goal is to choose $\epsilon$ such that $\Delta K(\tau_{\rm Th})$, see the circles in Fig.~\ref{figS2}, is larger than the noise.
We use $\epsilon=0.08$ for all results shown in the paper.
Such choice of $\epsilon$ gives rise to robustness of our results for different values of the widths $\eta$ of the spectral filter, and the agreement between results for $K(\tau)$ and $K_{\rm c}(\tau)$, see also Appendix~\ref{sec:robustness}. \\
(iii) The values of $\tau_{\rm Th}$ extracted from steps (i)-(ii) slightly underestimate the realistic $\tau_{\rm Th}$.
This is illustrated in Fig.~\ref{figS2} for the $J_1$-$J_2$ model at $L=18$ for different disorder strengths $W= 1, 3, 4$ and different widths $\eta = 0.1, 0.3, 0.5$ of the Gaussian filter $\rho(\varepsilon)$, for both the unconnected SFF [Figs.~\ref{figS2}(a)-\ref{figS2}(c)] and the connected SFF [Figs.~\ref{figS2}(d)-\ref{figS2}(f)].
We argue that a more accurate estimate of $\tau_{\rm Th}$ is obtained by requiring $\Delta K(\tau_{\rm Th}) = 10^{-2}$ in systems under investigation, see the horizontal lines in Fig.~\ref{figS2}.
This provides a compromise between setting $\Delta K$ as low as possible while being at the same time above the threshold set by finite-size fluctuations.
However, we do not perform the extrapolation of our results (circles in Fig.~\ref{figS2}) towards $\Delta K = 10^{-2}$ for each data set separately, but simply use it as an estimate of the global multiplicative factor being roughly equal to 4.
We hence set $\tau_{\rm Th} \to 4 \tau_{\rm Th}$ to get the final results shown in Figs.~\ref{fig2}-\ref{fig4} and Fig.~\ref{figS3}.
Note that the choice of the global multiplicative factor does not impact any of our main results.
Specifically, it only results in a constant shift of $g$, which alters neither the quality of the scaling solution nor the determination of the crossing point, and it is subtracted in the definition of the exponent $z$ in Eq.~(\ref{zW_extract}).

\section{Robustness of results} \label{sec:robustness}

\begin{figure}
\begin{center}
\includegraphics[width=0.80\columnwidth]{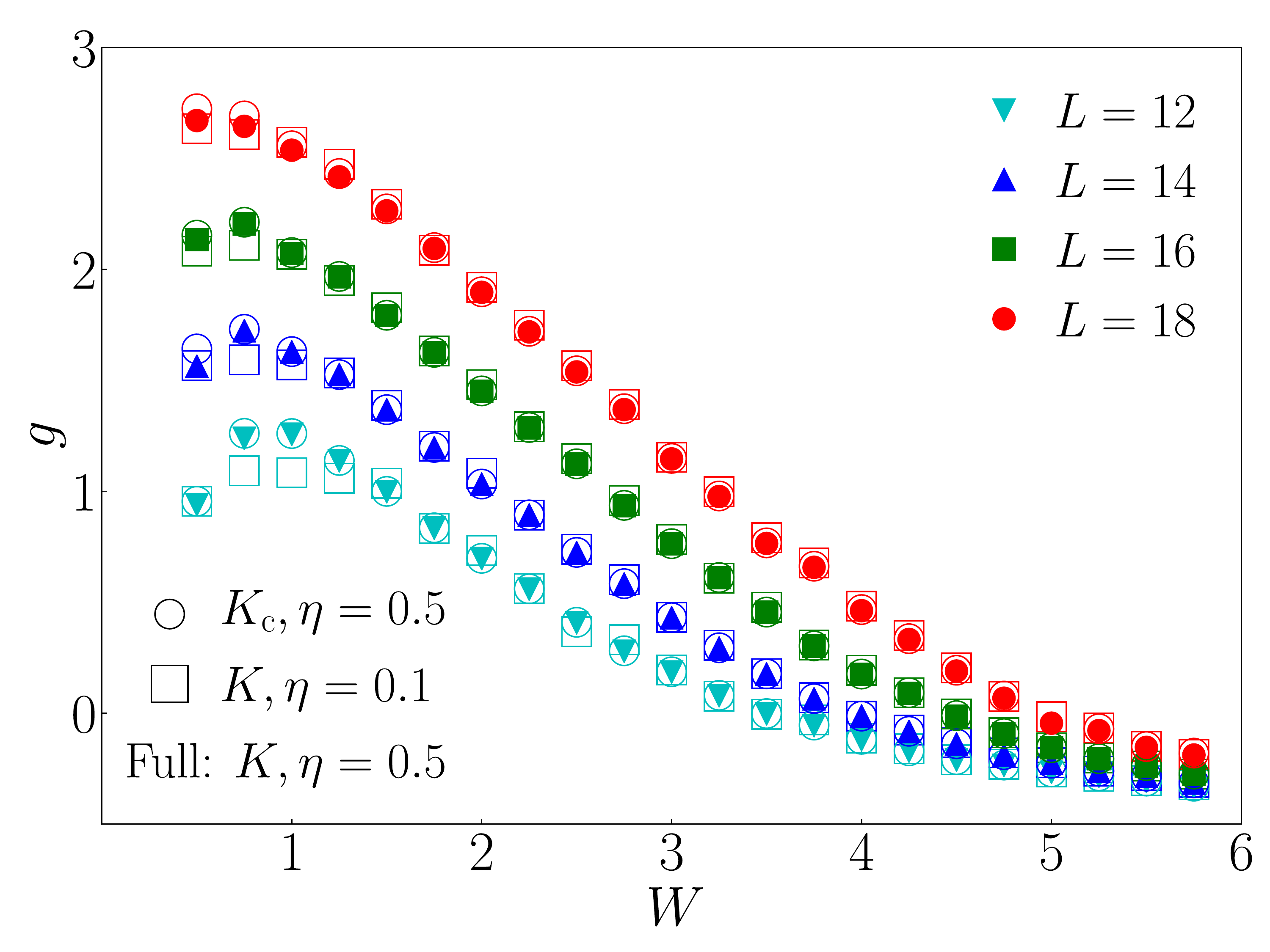}
\end{center}
\caption{{\it Comparison of the ergodicity indicator $g$ extracted from different implementations of the SFF $K(\tau)$.}
Results are shown as a function of the disorder strength $W$ for different system sizes $L$.
Filled symbols are identical to the results in Fig.~\ref{fig2}(a) in the main text.
Open squares are results using $t_{\rm Th}$ extracted from $K(\tau)$ with the Gaussian spectral filter of the width $\eta=0.1$.
Open circles are results using $t_{\rm Th}$ extracted from the connected SFF $K_{\rm c}(\tau)$ with $\eta=0.5$.
\label{figS3} }
\end{figure}

Finally, in Fig.~\ref{figS3} we show results for the ergodicity indicator $g=\log_{10}(t_{\rm H}/t_{\rm Th})$ in the $J_1$-$J_2$ model using the Thouless times $t_{\rm Th}$ extracted from three different implementations of the SFF:
(i) $K(\tau)$ with the Gaussian spectral filter with the width $\eta=0.5$, used also in the main text [filled symbols in Fig.~\ref{figS3}];
(ii) $K(\tau)$ with $\eta=0.1$ [open squares in Fig.~\ref{figS3}];
and (iii) the connected $K_{\rm c}(\tau)$ with $\eta=0.5$ [open circles in Fig.~\ref{figS3}].
The agreement between results using different implementations of the SFF confirms robustness of our protocol to extract the Thouless time.

\bibliographystyle{biblev1}
\bibliography{references_ergtransition}

\end{document}